\documentclass[prl,twocolumn]{revtex4}
\usepackage{dcolumn}
\usepackage{multirow}
\usepackage{graphicx}
\usepackage{amssymb}
\usepackage{bm}
\usepackage{hyperref}%for .pdf
\usepackage{epstopdf}%for .pdf
\usepackage{color}
\usepackage{mathrsfs}
\usepackage{amsmath,amssymb,amsthm}
\usepackage{rotating}
\usepackage{sverb, longtable}
\usepackage{subfigure}

\usepackage[graphicx]{realboxes}
\usepackage{adjustbox}
\begin{document}

\title{Questioning Cosmic Acceleration with DESI: The Big Stall of the Universe}
%Questioning Cosmic Acceleration with the Latest Observations
%Challenging Cosmic Acceleration with the Latest Observations
%The Fate of the Universe is Big Crunch in light of DESI DR2
\author{Deng Wang}
\email{dengwang@ific.uv.es}
\affiliation{Instituto de F\'{i}sica Corpuscular (CSIC-Universitat de Val\`{e}ncia), E-46980 Paterna, Spain} 

\begin{abstract}
One of the most important discoveries in modern cosmology is cosmic acceleration. However, we find that today's universe could decelerate in the statistically preferred Chevallier-Polarski-Linder (CPL) scenario over the $\Lambda$CDM model by cosmic microwave background, type Ia supernova and DESI's new measurements of baryon acoustic oscillations. Using various datasets, at a beyond $5\,\sigma$ confidence level, we demonstrate that the universe experiences a triple deceleration during its evolution and finally reaches the state of the ``Big Stall", which predicts that: (i) the universe suddenly comes to a halt in the distant future; (ii) its eventual destiny is dominated by dark matter rather than dark energy ; (iii) it ultimately retains an extremely small fraction of dark energy but exerts an extremely large pressure. Our findings profoundly challenge the established understanding of cosmic acceleration and enrich our comprehension of cosmic evolution.  

\end{abstract}
\maketitle

{\it Introduction.} The discovery of cosmic acceleration stands as one of the most profound revelations in modern cosmology, fundamentally altering our understanding of the universe's evolution and destiny. This phenomenon, which depicts the observed increase in the rate of the universe's expansion, has captivated scientists since its unexpected revelation. The implications of cosmic acceleration are vast, touching upon the very nature of space, time, and the fundamental forces that govern the cosmos.

Historically, the prevailing cosmological model suggested that the universe's expansion, initiated by the Big Bang, would gradually decelerate over time due to the gravitational attraction of matter. This view was grounded in Einstein's theory of general relativity and supported by observations of the early cosmic microwave background (CMB) radiation and the large-scale structure of the universe \cite{COBE:1992syq,Peebles1993,Mather:1993ij,Bennett:1996ce}. However, precise measurements of distant Type Ia supernovae (SN) challenged this perspective in the late 1990s. Two independent teams, the High-z Supernova Search Team \cite{SupernovaSearchTeam:1998fmf} and the Supernova Cosmology Project \cite{SupernovaCosmologyProject:1998vns}, revealed that these stellar explosions appeared fainter than expected. This inconsistency implied that cosmic expansion was not slowing down but rather accelerating.

The concept of cosmic acceleration introduced a new component into the cosmological framework: dark energy (DE). This enigmatic force, which permeates the fabric of space itself, is believed to drive the accelerated expansion of the universe. Nowadays,
the standard cosmological model is the so-called $\Lambda$CDM, where DE is usually represented by the cosmological constant ($\Lambda$). It also includes cold dark matter (CDM) and ordinary baryonic matter \cite{Bahcall:1999xn,Turner:1997npq,Carroll:2000fy}. While the $\Lambda$CDM scenario has been remarkably successful in explaining a wide range of cosmological observations, the nature of DE remains one of the most profound mysteries in contemporary physics \cite{Weinberg:1988cp,Peebles:2002gy,Padmanabhan:2002ji,Frieman:2008sn}.

Understanding cosmic acceleration is not merely an academic pursuit; it is essential for addressing fundamental questions about the universe's ultimate fate. If the acceleration continues unabated, the universe will expand indefinitely, leading to a cold and desolate future known as the ``heat death" of the universe \cite{Dyson:1979zz,Adams:1996xe}. In this scenario, the universe becomes increasingly empty as galaxies move farther apart, stars burn out, and the cosmos approaches a state of maximum entropy. This fate implies a universe devoid of usable energy, where no further work can be done, and all physical processes eventually cease.

Alternatively, if the nature of DE evolves over time, the universe's fate could be dramatically different. One possibility is the ``Big Rip'', where the accelerating expansion becomes so rapid that it overcomes the gravitational, electromagnetic, and nuclear forces holding matter together \cite{Caldwell:2003vq}. In this scenario, galaxies, stars, and even atoms are torn apart, leading to a catastrophic end of the universe. Another intriguing possibility is a cyclical model of the universe, where periods of expansion and contraction repeat indefinitely \cite{Steinhardt:2002ih,Penrose:2010zz}. In such models, the universe undergoes cycles of birth, expansion, contraction, and rebirth, potentially driven by the dynamics of DE or modifications to general relativity.

Furthermore, some theories propose that the accelerated expansion might slow down or even reverse in the distant future, leading to a ``Big Crunch'' where the universe collapses back into a singularity \cite{Barrow:1986}. Other speculative scenarios include the ``Big Freeze'', similar to the heat death but emphasizing the extreme cold and emptiness of the future universe, and the ``Vacuum Decay'', where a phase transition in the vacuum energy leads to a sudden and dramatic change in the properties of the universe \cite{Coleman:1980aw}.

During the past two decades, the CMB \cite{Planck:2018vyg,ACT:2025fju,SPT-3G:2022hvq,WMAP:2003elm,Planck:2013pxb}, baryon acoustic oscillations (BAO) \cite{SDSS:2005xqv,2dFGRS:2005yhx,Beutler:2011hx,eBOSS:2020yzd,deCarvalho:2017xye,eBOSS:2017cqx}, weak gravitational lensing \cite{Heymans:2012gg,Hildebrandt:2016iqg,Planck:2018lbu,DES:2017qwj}, galaxy clustering \cite{DES:2017myr,DES:2021wwk}, cluster counts \cite{DES:2025xii} and SN observations \cite{SupernovaSearchTeam:1998fmf,SupernovaCosmologyProject:1998vns} have further deepened our understanding of cosmic acceleration and confirmed the validity of $\Lambda$CDM. However, it confronts at least two challenges, i.e., the cosmological constant conundrum \cite{Weinberg:1988cp,Carroll:2000fy,Peebles:2002gy,Padmanabhan:2002ji} and the coincidence problem \cite{Steinhardt:1997}, while suffering from the emergent cosmic tensions of the Hubble constant ($H_0$) and the matter fluctuation amplitude ($S_8$) \cite{DiValentino:2020vhf,DiValentino:2020zio,DiValentino:2020vvd,Abdalla:2022yfr,DiValentino:2025sru}. Logically, these advancements compel theorists to study new physics beyond the $\Lambda$CDM model in order to resolve these issues (see \cite{Abdalla:2022yfr,DiValentino:2025sru} for reviews).

Recently, the new BAO measurements from the DESI data release two (DR2) \cite{DESI:2024mwx,DESI:2024uvr,DESI:2024lzq,DESI:2025zgx,DESI:2025zpo} inspired the explorations of new physics once again, especially, dynamical dark energy (DDE) \cite{DESI:2025fii}. Through the combinations of CMB, BAO and SN data, the DESI collaboration reported substantial evidences of DDE. However, in \cite{Wang:2025bkk}, we question the validity of such combinations, because there exist clear tensions among the datasets. Currently, a reliable approach is constraining the evolution of DE with each independent dataset and observe if there are statistical evidences of DDE. Interestingly, we find strong preferences of DDE over $\Lambda$CDM for each dataset used. This raises an even more interesting question: Whether cosmic acceleration still holds in such a preferred DDE universe? Employing various datasets, we find that current individual datasets cannot determine whether the present-day universe is accelerating or not.   

\begin{figure*}
	\centering
	\includegraphics[scale=0.5]{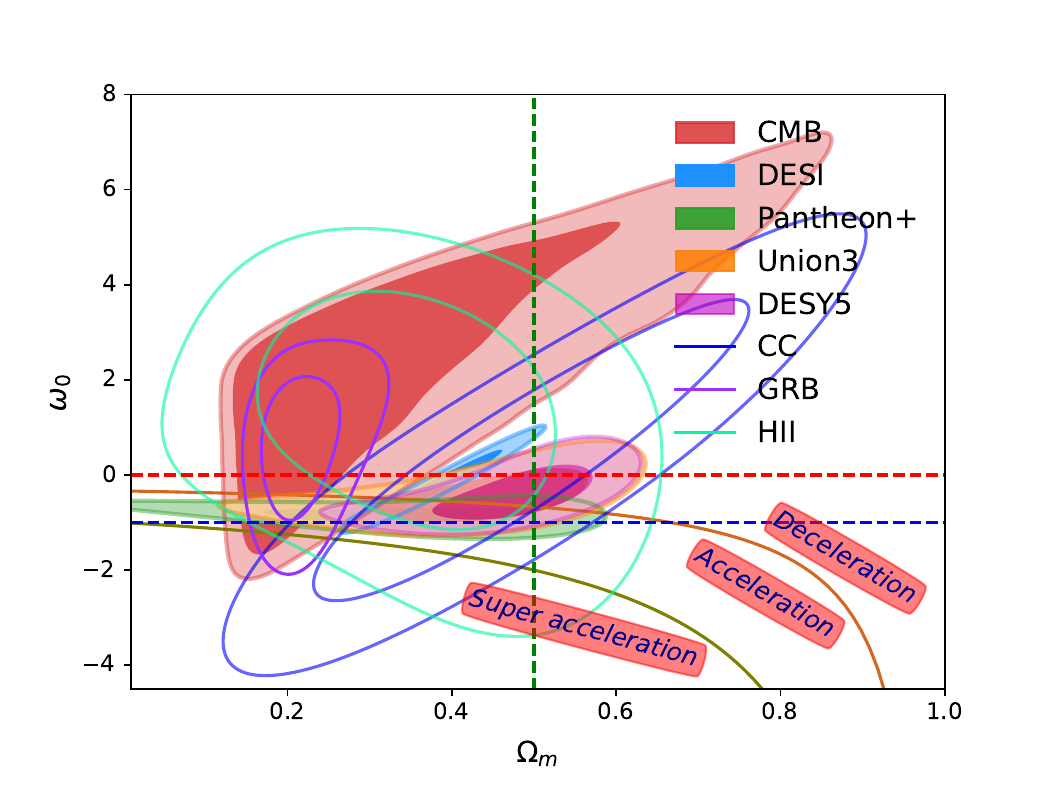}
	\includegraphics[scale=0.455]{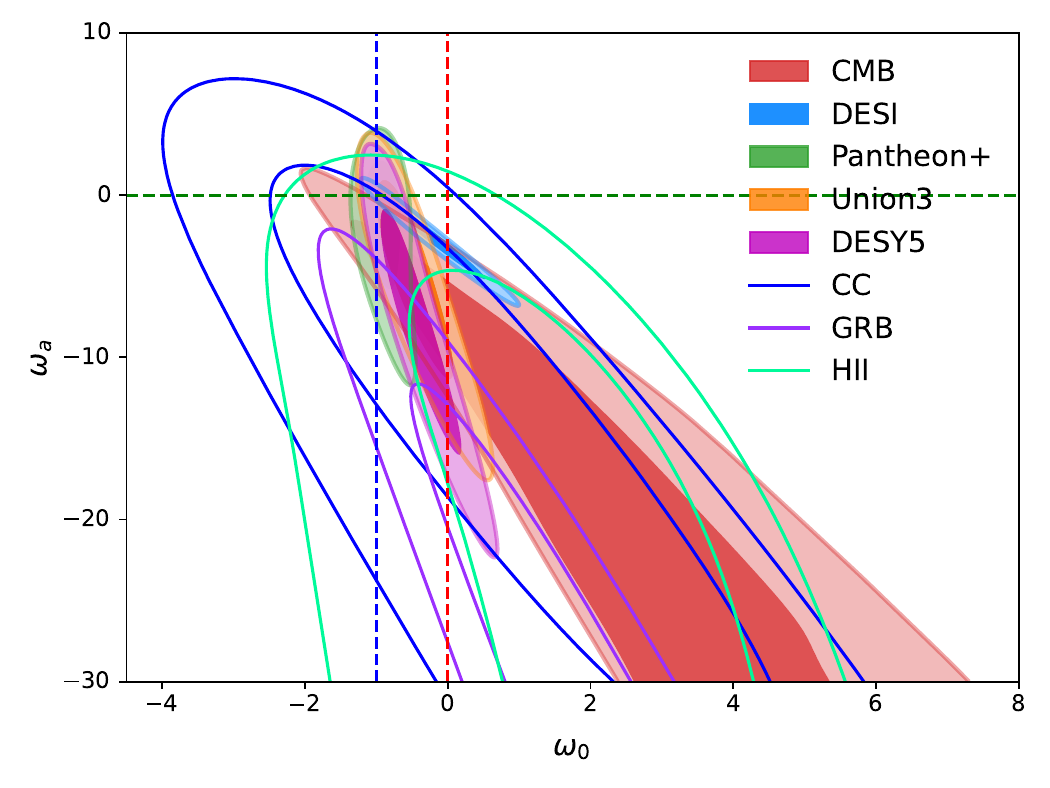}
	\caption{One-dimensional and two-dimensional posterior distributions of the parameter pairs $\Omega_m$-$\omega_0$ ({\it left}) and $\omega_0$-$\omega_a$ ({\it right}) from various datasets in the CPL model. The chocolate and olive solid lines are the boundaries of cosmic acceleration and super acceleration. The green dashed lines represents $\Omega_m=0.5$ ({\it left}) and $\omega_a=0$ ({\it right}). The red and blue dashed lines denote $\omega_0=0$ and $\omega_0=-1$, respectively.}\label{f1}
\end{figure*}

\begin{table*}[!t]
	\renewcommand\arraystretch{1.6}
	\begin{center}
		\caption{Mean values and $1\,\sigma$ (68.3\%) uncertainties of the present-day deceleration parameter $q_0$ from different datasets in the CPL model.}
		\setlength{\tabcolsep}{8mm}{
			\label{t1}
			\begin{tabular}{l c c c c c }
				\hline
				\hline
				Data & DESI & CMB & Pantheon+ & Union3 & DESY5     \\
				\hline 
				$q_0$ &  $0.34\pm 0.41$  & $2.83\pm 1.89$  & $-0.35^{+0.20}_{-0.21}$  & $0.12^{+0.29}_{-0.27}$  & $0.19\pm 0.28$  \\				
				\hline
				\hline
		\end{tabular}}
	\end{center}
\end{table*} 

%\begin{figure*}[h!]
%	\centering
%	\includegraphics[scale=0.3]{CA_evolution.pdf}
%	\caption{}\label{f2}
%\end{figure*}

\begin{figure*}
	\centering
	\includegraphics[width=1.02\textwidth, height=0.6\textheight]{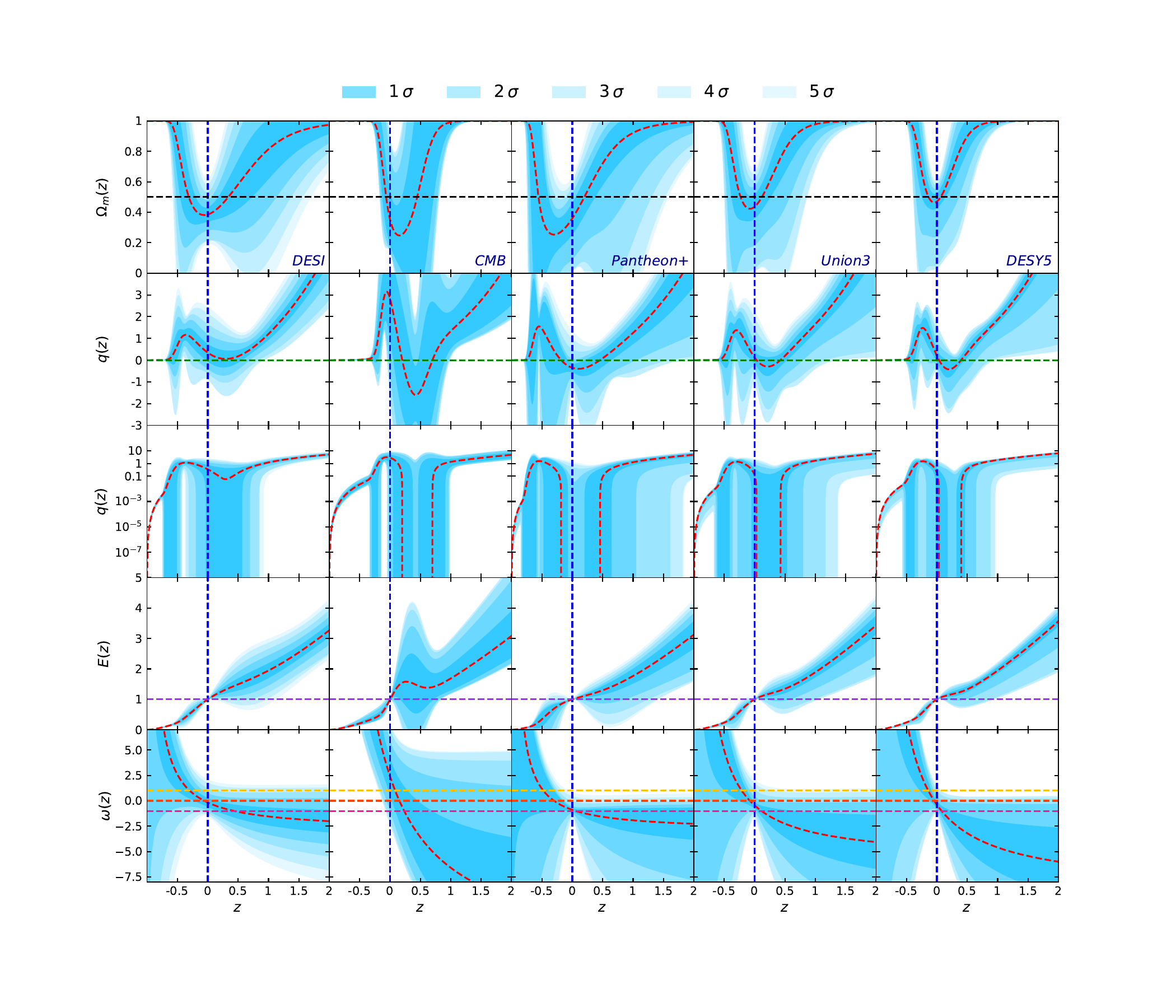}
	\caption{The $1\,\sigma-5\,\sigma$ regions of background quantities $\Omega_m(z)$, $q(z)$, $E(z)$ and $\omega(z)$ from five individual datasets in the CPL model. As a comparison, we plot $q(z)$ on a logarithmic scale in the third row. The dashed lines denote the mean values of each quantity (red), $z=0$ (blue), $\Omega_m=0.5$ (black), $q=0$ (green), $E=1$ (purple), $\omega_0=0$ (yellow), $\omega_0=0$ (orange) and $\omega_0=-1$ (pink), respectively.}\label{f2}
\end{figure*}

%\begin{widetext}
%	\begin{equation}
%	{\small E(z)=\left[\Omega_{m}(1+z)^3+(1-\Omega_{m})(1+z)^{3(1+\omega_0+\omega_a)}\mathrm{e}^{\frac{-3\omega_az}{1+z}}\right]^{\frac{1}{2}}}, \label{eq:ezcpl}
%	\end{equation}
%\end{widetext}

{\it Model.} In the framework of general relativity \cite{Einstein:1916vd}, considering a homogeneous and isotropic universe, the Friedmann equations read as $H^2=(8\pi G\rho)/3$ and $\ddot{a}/a=-4\pi G(\rho+3p)/3$, where $H$ is the cosmic expansion rate at a scale factor $a$ and $\rho$ and $p$ are the mean energy density and pressure of different species including baryons, dark matter (DM) and DE in the late universe. Combining two Friedmann equations, the normalized Hubble parameter $E(a)\equiv H(a)/H_0$ for a flat CPL universe \cite{Chevallier:2000qy,Linder:2002et} is shown as
\begin{equation}
 E(a)=\left[\Omega_{m}a^{-3}+(1-\Omega_{m})a^{-3(1+\omega_0+\omega_a)}\mathrm{e}^{3\omega_a(a-1)}\right]^{\frac{1}{2}}, \label{eq:ezcpl}
\end{equation}
where $\Omega_m$ is the matter fraction. It reduces to $\Lambda$CDM when $\omega_0=-1$ and $\omega_a=0$. In terms of the deceleration parameter $q\equiv-\ddot{a}/(aH^2)$, cosmic acceleration requires $\omega_0<1/[3(\Omega_m-1)]$ easily derived from the violation of strong energy condition $\rho+3p\geqslant0$, while a super acceleration requires $\omega_0<1/(\Omega_m-1)$ corresponding to the violation of null energy condition $\rho+p\geqslant0$.

{\it Data and methods.} We take 13 BAO measurements from DESI DR2 including the BGS, LRG1, LRG2, LRG3+ELG1, ELG2, QSO and Ly$\alpha$ samples at effective redshifts $z_{\rm eff}=0.295$,  0.51, 0.706, 0.934, 1.321, 1.484 and $2.33$, respectively \cite{DESI:2024mwx}. We use three calibrated SN datasets: (i) Pantheon+ including 1701 data points from 18 different surveys in $z\in[0.00122, 2.26137]$ \cite{Brout:2022vxf}; (ii) Union3 with 22 spline-interpolated data points derived by 2087 SN from 24 different surveys in $z\in[0.05, 2.26]$ \cite{Rubin:2023ovl}; (iii) DESY5 consisting of 1735 effective data points in $z\in[0.025, 1.130]$ \cite{DES:2024jxu}.
We employ the Planck 2018 high-$\ell$ \texttt{plik} temperature (TT) likelihood at multipoles  $30\leqslant\ell\leqslant2508$, polarization (EE) and their cross-correlation (TE) data at $30\leqslant\ell\leqslant1996$, and the low-$\ell$ TT \texttt{Commander} and \texttt{SimAll} EE likelihoods at $2\leqslant\ell\leqslant29$ \cite{Planck:2019nip}. We use conservatively the Planck lensing likelihood \cite{Planck:2018lbu} from \texttt{SMICA} maps at $8\leqslant\ell \leqslant400$. We also consider three complementary distance probes: 32 cosmic chronometers (CC) in $z\in[0.07, 1.96]$ \cite{Moresco:2016mzx}, 193 gamma ray bursts (GRB) in $z\in[0.03351, 8.1]$ \cite{Amati:2018tso} and 156 HII galaxies measurements $z\in[10^{-5}, 2.315]$ \cite{Terlevich:2015toa}.  

We use the Boltzmann code \texttt{CAMB} \cite{Lewis:1999bs} to compute the background evolution and theoretical power spectra of the universe. To perform the Bayesian analysis, we take the Monte Carlo Markov Chain (MCMC) method to infer the posterior distributions of model parameters using \texttt{Cobaya} \cite{Torrado:2020dgo}. We assess the convergence of MCMC chains via the Gelman-Rubin criterion $R-1\lesssim 0.01$ \cite{Gelman:1992zz} and analyze them using \texttt{Getdist} \cite{Lewis:2019xzd}.

We take the uniform priors for free parameters: the baryon fraction $\Omega_bh^2 \in [0.005, 0.1]$, cold DM fraction $\Omega_ch^2 \in [0.001, 0.99]$, acoustic angular scale at the recombination epoch $100\theta_{\rm MC} \in [0.5, 10]$, scalar spectral index $n_s \in [0.8, 1.2]$, amplitude of the primordial scalar power spectrum $\ln(10^{10}A_s) \in [2, 4]$, optical depth $\tau \in [0.01, 0.8]$, present-day DE EoS $\omega_0 \in [-15, 20]$ and amplitude of DE evolution $\omega_a \in [-30, 10]$. To produce a matter-dominated era at high redshifts, we impose the condition $\omega_0 + \omega_a < 0$ in the Bayesian analysis. The reason why we use such wide priors for ($\omega_0$, $\omega_a$) is that a large enough parameter space can completely exhibit the constraining power of DESI BAO data.

%and $H_0r_d \in [2946, 14730]$ km/s (for DESI only)

{\it Deceleration preferred by data.} Although the Planck collaboration has given a $2\,\sigma$ evidence of DDE \cite{Planck:2018vyg}, enough attention is not paid to it due to very loose constraints on ($\omega_0$, $\omega_a$) and the reason that combining previous BAO and SN with CMB data gives no deviation from $\Lambda$CDM \cite{Planck:2018vyg}. In light of the fact that new DESI BAO and SN data independently prefer DDE over $\Lambda$CDM \cite{Wang:2025bkk}, the CMB DDE evidence should be given sufficient attention. In Fig.\ref{f1}, we present the parameter spaces of ($\Omega_m$, $\omega_0$) from eight independent datasets including CC, GRB and HII observations, and find that, except for Pantheon+ that give a $1.75\,\sigma$ hint of cosmic acceleration, most of the parameter space from each dataset supports cosmic deceleration (see mean values in Tab.\ref{t1}). Specifically, CMB provides a $1.5\,\sigma$ evidence for deceleration, while DESI, Union3 and DESY5 gives $\sim 1\,\sigma$, $0.43\,\sigma$ and $0.68\,\sigma$ clues, respectively. To demonstrate the stability of our findings, using eight datasets, we show constraints on ($\omega_0$, $\omega_a$) in Fig.\ref{f1}. It is easy to see that all the datasets prefer the DDE region of $\omega_0>-1$ and $\omega_a<0$. Note that the CPL model is statistically favored over $\Lambda$CDM \cite{Wang:2025bkk} for these datasets. Therefore, we have verified that the universe could possibly undergo a present-day deceleration in such a preferred DDE universe by data. Interestingly, all the datasets cannot rule out $\omega_0>0$, indicating the possibility that the pressure of DE is positive. 

Furthermore, employing the error propagation, we derive the redshift evolution of background quantities $\Omega_m(z)$, $q(z)$, $E(z)$ and $\omega(z)$ from different datasets in the CPL model (see Fig.\ref{f2}). Overall, all the datasets support a triple deceleration at the $5\,\sigma$ confidence level, although the redshift ranges of deceleration are slightly different. It is a phenomenon similar to the staged acceleration of a rocket, but in the form of staged deceleration. One phase of deceleration occurred in the past, while the remaining two phases are anticipated to occur in the future. Unfortunately, we cannot determine if today's universe is accelerating with a high significance.

%{\it The Big Stall of the universe.}

{\it The Big Stall.} The $\Lambda$CDM model suggests a future where the universe becomes increasingly cold, dark, and empty, with the accelerating expansion driven by DE leading to the ultimate isolation and decay of all structures. However, this is not the case in the CPL model. In Fig.\ref{f2}, we find that all five datasets do not exclude the possibility of present-day matter domination, although DESI DR2 gives a $2.35\,\sigma$ evidence of $\Omega_m<0.5$. Based on five individual datasets, we confirm the following cosmic expansion history: After a long-term deceleration sourcing from the matter domination, DE starts to become comparable in magnitude to matter and mitigates the rate of deceleration of the universe. However, we cannot ensure whether the universe is accelerating now. Then, the universe will experience the second deceleration when the DE fraction significantly decreases and the universe becomes matter-dominated in the future. As the deceleration of the universe continues decreasing, it will enter the second phase where it is uncertain whether acceleration will resume. Subsequently, when the matter fraction is large enough, the universe will experience continuous deceleration until its acceleration ultimately reaches zero at the end of its existence. Interestingly, throughout this process, the rate of cosmic expansion continuously decreases and also reaches zero at the end of the universe. Conversely, the DE EoS continue increasing and finally become a positive infinity.
This indicates that the ultimate fate of the universe is completely stationary with $q(z\rightarrow-1)\rightarrow0$ and $E(z\rightarrow-1)\rightarrow0$ and matter-dominated. In the meanwhile, the universe will end up with an extremely small DE fraction but an extremely large DE pressure. We call this fate as the ``Big Stall''. Using the information of the baryon fraction from CMB \cite{Planck:2018vyg} or BBN \cite{Cooke:2017cwo}, the universe is DM-dominated at its end. Note that all the datasets roughly reveal that DE exhibits a phantom-crossing behavior at $2\,\sigma$ level, i.e., $\omega_0<-1$ in the past and $\omega_0>-1$ (and $\omega_0>0$) in the future.    

{\it Concluding remarks.} Currently, we cannot determine whether the universe is accelerating and even CMB, DESI DR2, Union3 and DESY5 data slightly prefer cosmic deceleration. This is different from the claim made by two SN teams who confirm cosmic acceleration under $\Lambda$. Because all the individual datasets favor the CPL DDE region of $\omega_0>-1$ and $\omega_a<0$ and prefer the CPL scenario over $\Lambda$CDM by a statistical comparison \cite{Wang:2025bkk}, our findings are logically reasonable. So far, there is no any independent dataset that can demonstrate the existence of today's cosmic acceleration in such a preferred universe. Based on this concern, the important contribution from two SN teams should be discovering the existence of dark energy not cosmic acceleration \cite{SupernovaSearchTeam:1998fmf,SupernovaCosmologyProject:1998vns}.  

The combinations of CMB, DESI and SN data give a similar fate of the universe to individual datasets. The main difference is that they strongly support today's cosmic acceleration or DE domination. However, their results are problematic \cite{DESI:2024mwx,DESI:2025zgx}, since they suffer from tensions among the datasets. For completeness, we show their background quantities in the supplementary material.

It is worth noting that all the datasets prefer $\omega(a)>1$ in the future, regardless of which dataset is used. This indicates that DE will behave like a ``super-stiff'' fluid, where the pressure is greater than its energy density. When $\omega(a)>1$ (see Fig.\ref{f2}), the sound speed of DE would be greater than the speed of light $c$. However, this contradicts the principle of special relativity \cite{Einstein:1905vqn}, which states that no information or energy can travel faster than the speed of light. 

In the CPL model, current observations give a beyond $5\,\sigma$ evidence of the Big Stall with a triple deceleration for the fate of the universe. Completely different from the Big Freeze and the Big Rip, the Big Stall predicts the following novel properties about the fate of the universe: (i) continuously decelerate; (ii) suddenly halt at a distant future; (iii) DM-dominated; (iv) extremely small DE fraction with extremely large pressure and EoS; (v) stars and galaxies could still be active; (vi) galaxies are not isolated;  (vii) no black hole era from the Big Freeze; (viii) it will not reaches a state of maximum entropy.

{\it Acknowledgements.} DW is supported by the CDEIGENT fellowship of Consejo Superior de Investigaciones Científicas (CSIC).

\appendix

% Create a title for the supplementary material
\onecolumngrid
%\section{\large Appendix}
\section{\large Supplementary Material for ``Questioning Cosmic Acceleration with DESI: The Big Stall of the Universe''}
\twocolumngrid

\onecolumngrid
{$\,\,\,\,$} In this supplementary material, we exhibit the redshift evolution of background quantities $\Omega_m(z)$, $q(z)$, $E(z)$ and $\omega(z)$ from four data combinations including CMB+DESI, CMB+DESI+Pantheon+, CMB+DESI+Union3 and CMB+DESI+DESY5 in the CPL model (see Fig.\ref{fs1}). 

{$\,\,\,\,$} Overall, these combinations provide tighter constraints on each background quantity due to the degeneracy breaking. All the combinations support today's universe is accelerating and dominated by dark energy, while they give stronger evidences of $\omega_0>-1$ (and $\omega_0>0$) in the distant future. Very interestingly, same as individual datasets, all these combinations reveal, a beyond $5\,\sigma$ confidence level, that the universe experiences a triple deceleration during its evolution and finally reaches the state of the ``Big Stall". Therefore, the beyond $5\,\sigma$ signal of the universe with a triple deceleration and the Big Stall does not depend on which datasets are used. 

\begin{figure}[h!]
	\setcounter{figure}{0}
	\centering
	\includegraphics[width=1.02\textwidth, height=0.6\textheight]{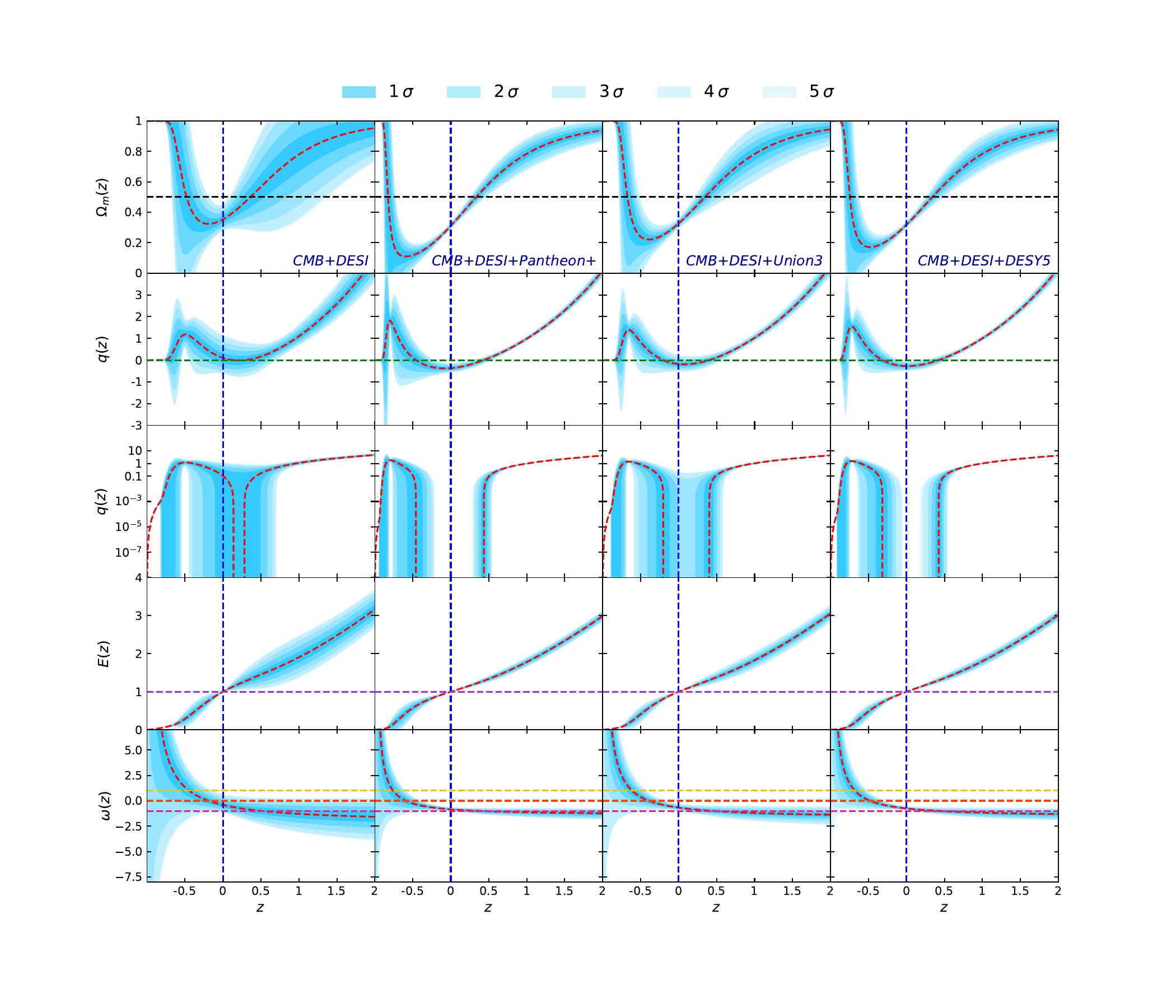}
	\caption{The $1\,\sigma-5\,\sigma$ regions of background quantities $\Omega_m(z)$, $q(z)$, $E(z)$ and $\omega(z)$ from different data combinations in the CPL model. As a comparison, we plot $q(z)$ on a logarithmic scale in the third row. The dashed lines denote the mean values of each quantity (red), $z=0$ (blue), $\Omega_m=0.5$ (black), $q=0$ (green), $E=1$ (purple), $\omega_0=0$ (yellow), $\omega_0=0$ (orange) and $\omega_0=-1$ (pink), respectively.}\label{fs1}
\end{figure}


\begin{thebibliography}{99}
%\cite{COBE:1992syq}
\bibitem{COBE:1992syq}
G.~F.~Smoot \textit{et al.} [COBE],
``Structure in the COBE differential microwave radiometer first year maps,''
Astrophys. J. Lett. \textbf{396}, L1-L5 (1992).
	
%\cite{Peebles1993}
\bibitem{Peebles1993}
P. J. E. Peebles, "Principles of Physical Cosmology." Princeton University Press (1993).

%\cite{Mather:1993ij}
\bibitem{Mather:1993ij}
J.~C.~Mather \textit{et al.},
``Measurement of the Cosmic Microwave Background spectrum by the COBE FIRAS instrument,''
Astrophys. J. \textbf{420}, 439-444 (1994).

%\cite{Bennett:1996ce}
\bibitem{Bennett:1996ce}
C.~L.~Bennett \textit{et al.},
``Four year COBE DMR cosmic microwave background observations: Maps and basic results,''
Astrophys. J. Lett. \textbf{464}, L1-L4 (1996).

%\cite{SupernovaSearchTeam:1998fmf}
\bibitem{SupernovaSearchTeam:1998fmf}
A.~G.~Riess \textit{et al.} [Supernova Search Team],
``Observational evidence from supernovae for an accelerating universe and a cosmological constant,''
Astron. J. \textbf{116}, 1009-1038 (1998).

%\cite{SupernovaCosmologyProject:1998vns}
\bibitem{SupernovaCosmologyProject:1998vns}
S.~Perlmutter \textit{et al.} [Supernova Cosmology Project],
``Measurements of $\Omega$ and $\Lambda$ from 42 high redshift supernovae,''
Astrophys. J. \textbf{517}, 565-586 (1999).

%\cite{Bahcall:1999xn}
\bibitem{Bahcall:1999xn}
N.~A.~Bahcall, J.~P.~Ostriker, S.~Perlmutter and P.~J.~Steinhardt,
``The Cosmic triangle: Assessing the state of the universe,''
Science \textbf{284}, 1481-1488 (1999).

%\cite{Turner:1997npq}
\bibitem{Turner:1997npq}
M.~S.~Turner and M.~J.~White,
``CDM models with a smooth component,''
Phys. Rev. D \textbf{56}, no.8, R4439 (1997).

%\cite{Carroll:2000fy}
\bibitem{Carroll:2000fy}
S.~M.~Carroll,
``The Cosmological constant,''
Living Rev. Rel. \textbf{4}, 1 (2001).

%\cite{Weinberg:1988cp}
\bibitem{Weinberg:1988cp} 
S.~Weinberg,
``The Cosmological Constant Problem,''
Rev.\ Mod.\ Phys.\  {\bf 61}, 1 (1989).

%\cite{Peebles:2002gy}
\bibitem{Peebles:2002gy}
P.~J.~E.~Peebles and B.~Ratra,
``The Cosmological Constant and Dark Energy,''
Rev. Mod. Phys. \textbf{75}, 559-606 (2003).

%\cite{Padmanabhan:2002ji}
\bibitem{Padmanabhan:2002ji}
T.~Padmanabhan,
``Cosmological constant: The Weight of the vacuum,''
Phys. Rept. \textbf{380}, 235-320 (2003).

%\cite{Frieman:2008sn}
\bibitem{Frieman:2008sn}
J.~Frieman, M.~Turner and D.~Huterer,
``Dark Energy and the Accelerating Universe,''
Ann. Rev. Astron. Astrophys. \textbf{46}, 385-432 (2008).

%\cite{Dyson:1979zz}
\bibitem{Dyson:1979zz}
F.~J.~Dyson,
``Time without end: Physics and biology in an open universe,''
Rev. Mod. Phys. \textbf{51}, 447-460 (1979).

%\cite{Adams:1996xe}
\bibitem{Adams:1996xe}
F.~C.~Adams and G.~Laughlin,
``A Dying universe: The Long term fate and evolution of astrophysical objects,''
Rev. Mod. Phys. \textbf{69}, 337-372 (1997).

%\cite{Caldwell:2003vq}
\bibitem{Caldwell:2003vq}
R.~R.~Caldwell, M.~Kamionkowski and N.~N.~Weinberg,
``Phantom energy and cosmic doomsday,''
Phys. Rev. Lett. \textbf{91}, 071301 (2003).

%\cite{Steinhardt:2002ih}
\bibitem{Steinhardt:2002ih}
P.~J.~Steinhardt, N.~Turok and N.~Turok,
``A Cyclic model of the universe,''
Science \textbf{296}, 1436-1439 (2002).

%\cite{Penrose:2010zz}
\bibitem{Penrose:2010zz}
R.~Penrose,
``The basic ideas of conformal cyclic cosmology,''
AIP Conf. Proc. \textbf{1446}, no.1, 233-243 (2012).

%\cite{Barrow:1986}
\bibitem{Barrow:1986}
J. D. Barrow and F. J. Tipler, "The Anthropic Cosmological Principle." Oxford University Press (1986).

%\cite{Coleman:1980aw}
\bibitem{Coleman:1980aw}
S.~R.~Coleman and F.~De Luccia,
``Gravitational Effects on and of Vacuum Decay,''
Phys. Rev. D \textbf{21}, 3305 (1980).


%\cite{Planck:2018vyg}
\bibitem{Planck:2018vyg}
N.~Aghanim \textit{et al.} [Planck],
``Planck 2018 results. VI. Cosmological parameters,''
Astron. Astrophys. \textbf{641}, A6 (2020)
[erratum: Astron. Astrophys. \textbf{652}, C4 (2021)].

%\cite{ACT:2025fju}
\bibitem{ACT:2025fju}
T.~Louis \textit{et al.} [ACT],
``The Atacama Cosmology Telescope: DR6 Power Spectra, Likelihoods and $\Lambda$CDM Parameters,''
[arXiv:2503.14452 [astro-ph.CO]].

%\cite{SPT-3G:2022hvq}
\bibitem{SPT-3G:2022hvq}
L.~Balkenhol \textit{et al.} [SPT-3G],
``Measurement of the CMB temperature power spectrum and constraints on cosmology from the SPT-3G 2018 TT, TE, and EE dataset,''
Phys. Rev. D \textbf{108}, no.2, 023510 (2023).


%\cite{WMAP:2003elm}
\bibitem{WMAP:2003elm}
D.~N.~Spergel \textit{et al.} [WMAP],
``First year Wilkinson Microwave Anisotropy Probe (WMAP) observations: Determination of cosmological parameters,''
Astrophys. J. Suppl. \textbf{148}, 175-194 (2003).

%\cite{Planck:2013pxb}
\bibitem{Planck:2013pxb}
P.~A.~R.~Ade \textit{et al.} [Planck],
``Planck 2013 results. XVI. Cosmological parameters,''
Astron. Astrophys. \textbf{571}, A16 (2014).

%\cite{SDSS:2005xqv}
\bibitem{SDSS:2005xqv}
D.~J.~Eisenstein \textit{et al.} [SDSS],
``Detection of the Baryon Acoustic Peak in the Large-Scale Correlation Function of SDSS Luminous Red Galaxies,''
Astrophys. J. \textbf{633}, 560-574 (2005).

%\cite{2dFGRS:2005yhx}
\bibitem{2dFGRS:2005yhx}
S.~Cole \textit{et al.} [2dFGRS],
``The 2dF Galaxy Redshift Survey: Power-spectrum analysis of the final dataset and cosmological implications,''
Mon. Not. Roy. Astron. Soc. \textbf{362}, 505-534 (2005).

%\cite{Beutler:2011hx}
\bibitem{Beutler:2011hx}
F.~Beutler \textit{et al.} [6dF],
``The 6dF Galaxy Survey: Baryon Acoustic Oscillations and the Local Hubble Constant,''
Mon. Not. Roy. Astron. Soc. \textbf{416}, 3017-3032 (2011)


%\cite{eBOSS:2020yzd}
\bibitem{eBOSS:2020yzd}
S.~Alam \textit{et al.} [eBOSS],
``Completed SDSS-IV extended Baryon Oscillation Spectroscopic Survey: Cosmological implications from two decades of spectroscopic surveys at the Apache Point Observatory,''
Phys. Rev. D \textbf{103}, no.8, 083533 (2021).

%\cite{deCarvalho:2017xye}
\bibitem{deCarvalho:2017xye}
E.~de Carvalho, A.~Bernui, G.~C.~Carvalho, C.~P.~Novaes and H.~S.~Xavier,
``Angular Baryon Acoustic Oscillation measure at $z=2.225$ from the SDSS quasar survey,''
JCAP \textbf{04}, 064 (2018).

%\cite{eBOSS:2017cqx}
\bibitem{eBOSS:2017cqx}
M.~Ata \textit{et al.} [eBOSS],
``The clustering of the SDSS-IV extended Baryon Oscillation Spectroscopic Survey DR14 quasar sample: first measurement of baryon acoustic oscillations between redshift 0.8 and 2.2,''
Mon. Not. Roy. Astron. Soc. \textbf{473}, no.4, 4773-4794 (2018).


%\cite{Heymans:2012gg}
\bibitem{Heymans:2012gg}
C.~Heymans, \textit{et al.} [CFHTLenS],
``CFHTLenS: The Canada-France-Hawaii Telescope Lensing Survey,''
Mon. Not. Roy. Astron. Soc. \textbf{427}, 146 (2012).

%\cite{Hildebrandt:2016iqg}
\bibitem{Hildebrandt:2016iqg}
H.~Hildebrandt \textit{et al.} [KiDS],
``KiDS-450: Cosmological parameter constraints from tomographic weak gravitational lensing,''
Mon. Not. Roy. Astron. Soc. \textbf{465}, 1454 (2017).

%\cite{Planck:2018lbu}
\bibitem{Planck:2018lbu}
N.~Aghanim \textit{et al.} [Planck],
``Planck 2018 results. VIII. Gravitational lensing,''
Astron. Astrophys. \textbf{641}, A8 (2020).

%\cite{DES:2017qwj}
\bibitem{DES:2017qwj}
M.~A.~Troxel \textit{et al.} [DES],
``Dark Energy Survey Year 1 results: Cosmological constraints from cosmic shear,''
Phys. Rev. D \textbf{98}, no.4, 043528 (2018).

%\cite{DES:2017myr}
\bibitem{DES:2017myr}
T.~M.~C.~Abbott \textit{et al.} [DES],
``Dark Energy Survey year 1 results: Cosmological constraints from galaxy clustering and weak lensing,''
Phys. Rev. D \textbf{98}, no.4, 043526 (2018).


%\cite{DES:2021wwk}
\bibitem{DES:2021wwk}
T.~M.~C.~Abbott \textit{et al.} [DES],
``Dark Energy Survey Year 3 results: Cosmological constraints from galaxy clustering and weak lensing,''
Phys. Rev. D \textbf{105}, no.2, 023520 (2022).


%\cite{DES:2025xii}
\bibitem{DES:2025xii}
T.~M.~C.~Abbott \textit{et al.} [DES],
``Dark Energy Survey Year 3 Results: Cosmological Constraints from Cluster Abundances, Weak Lensing, and Galaxy Clustering,''
[arXiv:2503.13632 [astro-ph.CO]].

%\cite{Steinhardt:1997}
\bibitem{Steinhardt:1997}
P. J. Steinhardt, "Critical Problems in Physics." Princeton University Press (1997).

%\cite{DiValentino:2020vhf}
\bibitem{DiValentino:2020vhf}
E.~Di Valentino \textit{et al.},
``Snowmass2021 - Letter of interest cosmology intertwined I: Perspectives for the next decade,''
Astropart. Phys. \textbf{131}, 102606 (2021).

%\cite{DiValentino:2020zio}
\bibitem{DiValentino:2020zio}
E.~Di Valentino \textit{et al.},
``Snowmass2021 - Letter of interest cosmology intertwined II: The hubble constant tension,''
Astropart. Phys. \textbf{131}, 102605 (2021).

%\cite{DiValentino:2020vvd}
\bibitem{DiValentino:2020vvd}
E.~Di Valentino \textit{et al.},
``Cosmology Intertwined III: $f \sigma_8$ and $S_8$,''
Astropart. Phys. \textbf{131}, 102604 (2021).

%\cite{Abdalla:2022yfr}
\bibitem{Abdalla:2022yfr}
E.~Abdalla \textit{et al.},
``Cosmology intertwined: A review of the particle physics, astrophysics, and cosmology associated with the cosmological tensions and anomalies,''
JHEAp \textbf{34}, 49-211 (2022).

%\cite{DiValentino:2025sru}
\bibitem{DiValentino:2025sru}
E.~Di Valentino \textit{et al.},
``The CosmoVerse White Paper: Addressing observational tensions in cosmology with systematics and fundamental physics,''
[arXiv:2504.01669 [astro-ph.CO]].

%\cite{DESI:2024mwx}
\bibitem{DESI:2024mwx}
A.~G.~Adame \textit{et al.} [DESI],
``DESI 2024 VI: Cosmological Constraints from the Measurements of Baryon Acoustic Oscillations,''
[arXiv:2404.03002 [astro-ph.CO]].

%\cite{DESI:2024uvr}
\bibitem{DESI:2024uvr}
A.~G.~Adame \textit{et al.} [DESI],
``DESI 2024 III: Baryon Acoustic Oscillations from Galaxies and Quasars,''
[arXiv:2404.03000 [astro-ph.CO]].

%\cite{DESI:2024lzq}
\bibitem{DESI:2024lzq}
A.~G.~Adame \textit{et al.} [DESI],
``DESI 2024 IV: Baryon Acoustic Oscillations from the Lyman Alpha Forest,''
[arXiv:2404.03001 [astro-ph.CO]].

%\cite{DESI:2025zgx}
\bibitem{DESI:2025zgx}
M.~Abdul Karim \textit{et al.} [DESI],
``DESI DR2 Results II: Measurements of Baryon Acoustic Oscillations and Cosmological Constraints,''
[arXiv:2503.14738 [astro-ph.CO]].


%\cite{DESI:2025zpo}
\bibitem{DESI:2025zpo}
M.~Abdul Karim \textit{et al.} [DESI],
``DESI DR2 Results I: Baryon Acoustic Oscillations from the Lyman Alpha Forest,''
[arXiv:2503.14739 [astro-ph.CO]].

%\cite{DESI:2025fii}
\bibitem{DESI:2025fii}
K.~Lodha \textit{et al.} [DESI],
``Extended Dark Energy analysis using DESI DR2 BAO measurements,''
[arXiv:2503.14743 [astro-ph.CO]].

%\cite{Wang:2025bkk}
\bibitem{Wang:2025bkk}
D.~Wang and D.~Mota,
``Did DESI DR2 truly reveal dynamical dark energy?''
[arXiv:2504.15222 [astro-ph.CO]].

%\cite{Einstein:1916vd}
\bibitem{Einstein:1916vd}
A.~Einstein,
``The foundation of the general theory of relativity.,''
Annalen Phys. \textbf{49}, no.7, 769-822 (1916).

%\cite{Chevallier:2000qy}
\bibitem{Chevallier:2000qy}
M.~Chevallier and D.~Polarski,
``Accelerating universes with scaling dark matter,''
Int. J. Mod. Phys. D \textbf{10}, 213-224 (2001).

%\cite{Linder:2002et}
\bibitem{Linder:2002et}
E.~V.~Linder,
``Exploring the expansion history of the universe,''
Phys. Rev. Lett. \textbf{90}, 091301 (2003).


%\cite{Brout:2022vxf}
\bibitem{Brout:2022vxf}
D.~Brout \textit{et al.},
``The Pantheon+ Analysis: Cosmological Constraints,''
Astrophys. J. \textbf{938}, no.2, 110 (2022).

%\cite{Rubin:2023ovl}
\bibitem{Rubin:2023ovl}
D.~Rubin \textit{et al.},
``Union Through UNITY: Cosmology with 2,000 SNe Using a Unified Bayesian Framework,''
[arXiv:2311.12098 [astro-ph.CO]].


%\cite{DES:2024jxu}
\bibitem{DES:2024jxu}
T.~M.~C.~Abbott \textit{et al.} [DES],
``The Dark Energy Survey: Cosmology Results with \ensuremath{\sim}1500 New High-redshift Type Ia Supernovae Using the Full 5 yr Data Set,''
Astrophys. J. Lett. \textbf{973}, no.1, L14 (2024).

%\cite{Planck:2019nip}
\bibitem{Planck:2019nip}
N.~Aghanim \textit{et al.} [Planck],
``Planck 2018 results. V. CMB power spectra and likelihoods,''
Astron. Astrophys. \textbf{641}, A5 (2020).



%\cite{Moresco:2016mzx}
\bibitem{Moresco:2016mzx}
M.~Moresco \textit{et al.},
``A 6\% measurement of the Hubble parameter at $z\sim0.45$: direct evidence of the epoch of cosmic re-acceleration,''
JCAP \textbf{05}, 014 (2016).

%\cite{Amati:2018tso}
\bibitem{Amati:2018tso}
L.~Amati, R.~D'Agostino, O.~Luongo, M.~Muccino and M.~Tantalo,
``Addressing the circularity problem in the $E_\text{p}-E_\text{iso}$ correlation of gamma-ray bursts,''
Mon. Not. Roy. Astron. Soc. \textbf{486}, no.1, L46-L51 (2019).

%\cite{Terlevich:2015toa}
\bibitem{Terlevich:2015toa}
R.~Terlevich \textit{et al.},
``On the road to precision cosmology with high-redshift H II galaxies,''
Mon. Not. Roy. Astron. Soc. \textbf{451}, no.3, 3001-3010 (2015).

%\cite{Lewis:1999bs}
\bibitem{Lewis:1999bs}
A.~Lewis, A.~Challinor and A.~Lasenby,
``Efficient computation of CMB anisotropies in closed FRW models,''
Astrophys. J. \textbf{538}, 473-476 (2000).

%\cite{Torrado:2020dgo}
\bibitem{Torrado:2020dgo}
J.~Torrado and A.~Lewis,
``Cobaya: Code for Bayesian Analysis of hierarchical physical models,''
JCAP \textbf{05}, 057 (2021).

%\cite{Gelman:1992zz}
\bibitem{Gelman:1992zz}
A.~Gelman and D.~B.~Rubin,
``Inference from Iterative Simulation Using Multiple Sequences,''
Statist. Sci. \textbf{7}, 457-472 (1992).

%\cite{Lewis:2019xzd}
\bibitem{Lewis:2019xzd}
A.~Lewis,
``GetDist: a Python package for analysing Monte Carlo samples,''
[arXiv:1910.13970 [astro-ph.IM]].

%\cite{Cooke:2017cwo}
\bibitem{Cooke:2017cwo}
R.~J.~Cooke, M.~Pettini and C.~C.~Steidel,
``One Percent Determination of the Primordial Deuterium Abundance,''
Astrophys. J. \textbf{855}, no.2, 102 (2018).

%\cite{Einstein:1905vqn}
\bibitem{Einstein:1905vqn}
A.~Einstein,
``Zur Elektrodynamik bewegter K\"orper,''
Annalen Phys. \textbf{322}, no.10, 891-921 (1905).

\clearpage
\end{thebibliography}
\end{document}